\documentclass[preprint]{elsarticle}

\usepackage{hyperref}
 \usepackage{graphicx}

\usepackage{amssymb}

\begin{document}

\begin{frontmatter}



\title{Tuner: a tool for designing and optimizing ion optical systems}


\author[NSCL]{D. Bazin}
\ead{bazin@nscl.msu.edu}

\address[NSCL]{National Superconducting Cyclotron Laboratory, Michigan State University, East Lansing, MI 48824-1321, USA}

\begin{abstract}
Designing and optimizing ion optical systems is often a complex and difficult task, which requires the use of computational tools to iterate and converge towards the desired characteristics and performances of the system. Very often these tools are not well adapted for exploring the numerous degrees of freedom, rendering the process long and tedious, as well as somewhat random due to the very large number of local minima typically found when looking for a particular optical solution. This paper presents a novel approach to finding the desired solution of an optical system, by providing the user with an instant feedback of the effects of changing parameters. The process of finding an approximate solution by manually adjusting parameters is greatly facilitated, at which point the final tune can be calculated by minimization according to a number of constraints.
\end{abstract}

\end{frontmatter}

\section{Introduction}
\label{introduction}
Historically, the design and optimization of ion optical systems has been performed by means of computer programs in which the precise specification of the system is entered and its optical properties, such as beam envelopes or transfer matrix coefficients, are obtained as output. Probably the most famous software of this kind is the program TRANSPORT \cite{Transport}, which is still widely used nowadays. These tools have been developed over many years, and although several of them have been modernized to follow the evolution of computer environments, they often retain legacy features of the early days. The lack of interactivity in particular, due to the mode of entering the description of the system via a file, makes the process of exploration and optimization very tedious.

For all but the most simple systems, finding an optical solution fulfilling a set of predefined constraints is equivalent to a multi-dimensional minimization problem. This type of problem often displays multiple local minima in which a minimization algorithm can easily get trapped, depending on the variational path or method used. This means practically that the solution found by the algorithm is often highly dependent on the initial conditions at the start of the minimization process. As a result, finding a configuration close enough to the desired solution becomes a crucial task, which cannot be performed by the machine. The program Tuner was designed to make this task more effective as well as more intuitive, by taking advantage of the computing power and graphic tools available today. Therefore, this program is intended as a design tool used to try out new ideas or explore new optical solutions of an existing design, rather than a simulation tool trying to calculate the characteristics of an optical system as accurately as possible. Several programs are already available for this last task, such as COSY Infinity \cite{COSY} for instance, that can be used to optimize and simulate a design once its gross characteristics have been laid out, from an accurate knowledge of the elements involved and the possibility to calculate high order terms of the transfer matrix.

The next section presents the calculations as well as the framework on which the program is based, followed by a description of how to perform the design and optimization processes of an optical system. Finally, some examples of existing systems and how to explore their optical possibilities are given, prior to the conclusion.

\section{Framework and calculations}
\label{framework}
\subsection{Framework and environment}
\begin{figure}
\begin{center}
\includegraphics[scale=0.6]{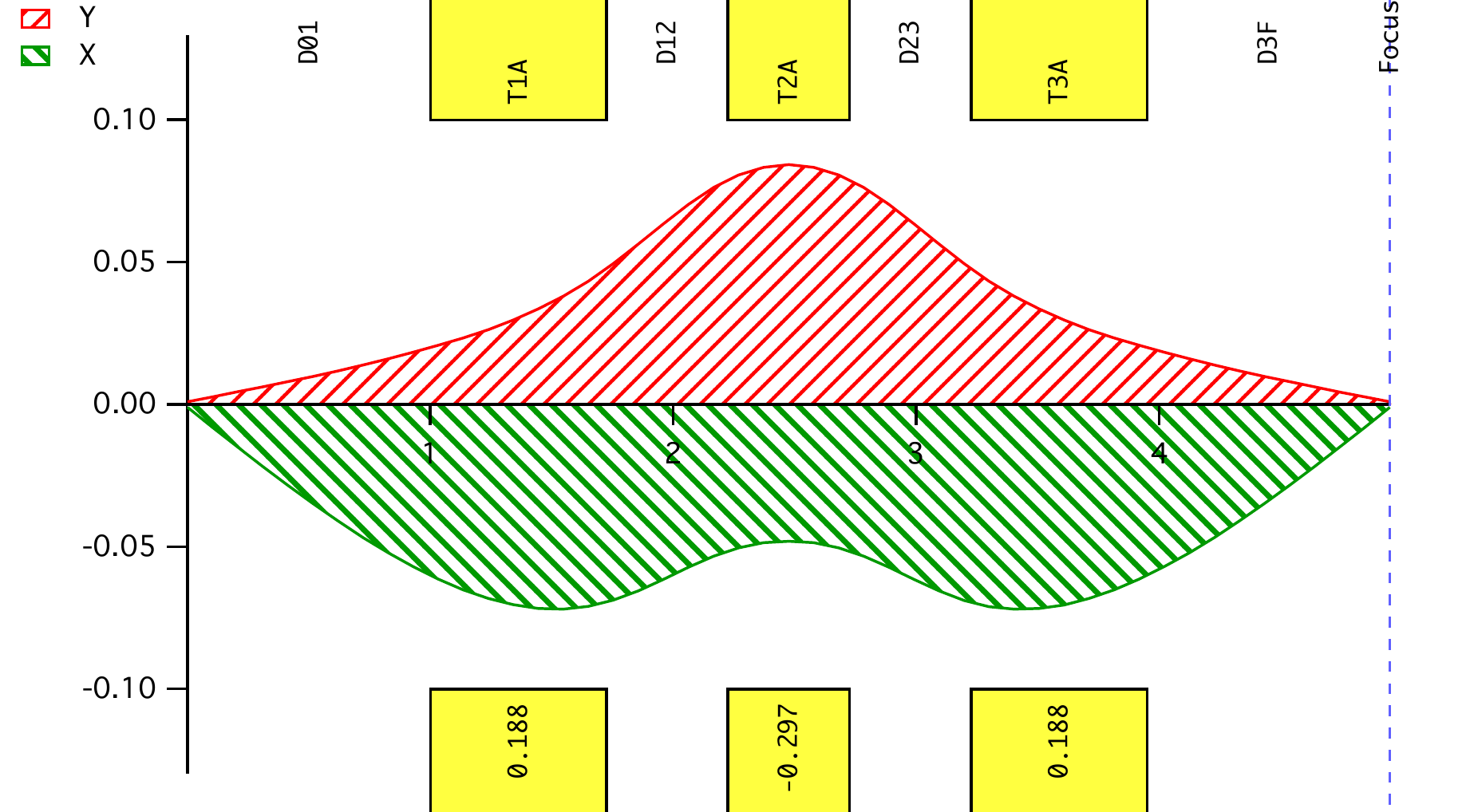}
\caption{Example of an envelope plot produced by Tuner. Following the convention in TRANSPORT, the vertical envelope is shown on top of the horizontal axis (slanted up pattern and red in online version), while the horizontal envelope is shown at the bottom (slanted down pattern and green in online version). The boxes (yellow in online version) indicate the position and bore of the quadrupoles, as well as their names and applied field at pole tip (in Tesla). The names of the drift spaces in between are also indicated, as is the location of the final focus of the system. All distances are in meters. Note that the actual envelope plots produced by the program do not have the slanted patterns, which are included in this figure only for additional clarity in the black-and-white printed version of this article.}
\label{envelopes}
\end{center}
\end{figure}

The program Tuner was developed and written within the framework of the data analysis and visualization software Igor Pro$\textsuperscript{\textregistered}$ (WaveMetrics, Inc., Oregon, USA) \cite{IgorPRO}. The choice of this particular environment is motivated by several features offered by this package. The first and foremost is the ability to create interactive graphics to display the characteristics of the system programmatically. Since all features and functions offered by Igor Pro$\textsuperscript{\textregistered}$ can be used to build customized macros, displaying the configuration and characteristics of the system can be done easily in a few lines of code. As an example, Fig. \ref{envelopes} shows a typical envelope plot obtained for a quadrupole triplet tuned in point-to-point focussing.
The graphics tools provided in Igor Pro$\textsuperscript{\textregistered}$ are well adapted to the generation of such a plot, not only because they allow to easily draw the envelopes themselves as well as the elements present in the system, but also because all graphs are dynamically updated whenever the underlying data is changed. This feature is essential to this application, to provide an instant visual feedback to the user while adjusting parameters.

Secondly, the library of mathematical and data processing functions provided in Igor Pro$\textsuperscript{\textregistered}$ greatly simplified and shortened the development time of this program. Of the numerous functions available, the Tuner program uses extensively the matrix algebra and manipulation functions, as well as the optimization and minimization algorithms. Since these functions are already compiled in machine code in Igor Pro$\textsuperscript{\textregistered}$, the speed of execution is the maximum possible. This means for instance that the envelope and ray plots can be recalculated instantly whenever any parameter of the system is changed, providing a "live" mode essential when looking for an approximate solution manually.

Thirdly, the possibility to create customized graphical user interfaces (GUI) means that users don't need to learn how to use Igor Pro$\textsuperscript{\textregistered}$ itself, because all necessary actions are available from the customized Tuner control panel. This feature gives great flexibility to tailor the interface for the particular needs of the application, and shortens the learning curve of the users.

Finally, the software Igor Pro$\textsuperscript{\textregistered}$ is available for both MacIntosh$\textsuperscript{\textregistered}$ and Windows$\textsuperscript{\textregistered}$ platforms, covering the majority of personal computers. Users are required to purchase a license in order to use this program, and academic pricing is available for scholars and students (see \href{http://www.wavemetrics.com}{www.wavemetrics.com}).

\subsection{Beam line calculations and supported elements}
The coordinate system used in Tuner is the usual 6-dimensional space defined as horizontal position $x$ and angle $a = x'$, and their vertical counterparts $y$ and $b = y'$, path length $\ell$ and fractional momentum deviation $d = \Delta p/p$. Although $a$ and $b$ are called angles, strictly speaking they are in fact the tangents of these angles, but for the small angles usually under consideration in optical systems the difference is negligible. The units of these coordinates are meters and radians for positions and angles respectively, meters for the path length and 'parts' or '1' for the fractional momentum deviation. This last unit simply means that a fractional momentum deviation of 1\% corresponds to 0.01 in parts.

At the design and early optimization stages of an optical system, only first order optics are usually considered, because they determine the gross characteristics of the system. The first order transfer matrix of the most common elements found in beam lines are well known \cite{Transport}. These include drift space, magnetic quadrupoles and dipoles, as well as solenoids. The program calculates the first order transfer and envelope (also called beam or sigma) matrices along the optical system, by multiplying the individual matrices of each element. More formally, the coordinate vector $V_B$ at location $B$ is obtained from the coordinate vector $V_A$ at location $A$ following the relation: 
$$V_B = R_{BA}V_A,$$
where the 6x6 transfer matrix $R_{BA}$ is obtained from the multiplication of all matrices from A to B: 
$$R_{BA} = R_n \cdots R_2 R_1,$$
assuming there are $n$ elements between $A$ and $B$. Similarly, the beam matrix $\sigma_B$ is obtained following the relation: 
$$\sigma_B = R_{BA} \sigma_A R^T_{BA}.$$
From the set of transfer and beam matrices rays and envelopes can be plotted along the optical system. In addition, particular locations along the beam line can be specified and are labeled Viewers. These Viewers are typically located at dispersive or focal planes, and allow the user to inspect the characteristics of the optical system, such as the transfer and envelope matrices, or the shapes of the beam ellipses. Finally, fitting constraints can be inserted anywhere in the system, and are used to define the desired properties of the optical system at various locations. Although not required, it is usually desirable to group fitting constraints close-by viewers, so that the properties of the system can be inspected upon attempting to satisfy the fitting conditions. Finally, the characteristics of the beam propagated into the system is always defined as the first element of a given system. The program allows the user to specify the sizes of the beam ellipse in all dimensions, as well as the value of the magnetic rigidity (denoted as $B\rho$) of the reference particle, and optionally the orientation of the beam ellipse in both horizontal and vertical planes. These parameters are denoted $r_{12}$ and $r_{34}$ for these two planes respectively, and correspond to the off-diagonal elements $(a/x)$ and $(b/y)$ of the beam matrix. In addition, it is possible to use the last beam matrix of a given system as the start of another. This feature gives the user the flexibility to divide a large system into smaller sub-systems, thereby easier to display and optimize, while retaining the full calculation.

\subsection{Varying parameters and fitting}
\label{varlim}
The design of an optical system can be thought of as an iterative process, during which the designer goes back and forth between altering the design and fitting the required constraints of the system. The most common parameters that are used to fit the constraints are the strengths of the lenses. However, and especially in the early phases of a design, it is often desirable to be able to vary other parameters, such as the length of quadrupoles or pole face angle of dipoles, in order to attain the desired characteristics of the system. The program Tuner allows the user to vary most of the parameters of the supported elements.

Fitting locations can be defined along the system, and are usually placed right after a set of constraints. These can be enabled or disabled, so that the elements for which varying parameters enter in the fitting function go only until an enabled  fit location is encountered. This feature can greatly reduce the CPU time spent in the fitting by limiting the fitting iteration to a partial section of beam line. Individual varying parameters and constraints can also be enabled or disabled to the same effect, but using the fitting locations is much more convenient and keeps the choices of varying parameters and constraints in the sections unaffected by the fit untouched.

The fitting function is formed from the sum of all enabled constraints, each of them being specified by a desired target value and a tolerance. The tolerance is used to specify the desired weight and directly affects the level of accuracy obtained in the fit. Both transfer and beam matrix elements can be used to define the constraints. The fitting function is similar to a $\chi^2$ minimization function and has the form:
$$\chi^2 = \sum_{i=1}^n \frac{(M_i - V_i)^2}{T_i^2},$$
where the $M_i$ and $V_i$ are the matrix elements and their desired target values respectively, and the $T_i$ are the tolerances.

Several fitting algorithms are available in Igor Pro$\textsuperscript{\textregistered}$, both to calculate the next step and the matrix of second derivatives (also called Hessian). The default algorithms are the line search and secant methods for the next step and second derivatives, respectively. Other possibilities are the dogleg and More-Hebdon algorithms for the next step determination, and the finite differences algorithm for the second derivatives. The fitting of multivariate functions is a notoriously difficult task in computing. Most problems arise when the function is either under-constrained or over-constrained. When under-constrained, the minimization function usually exhibits a large number of minima, which makes it difficult to find the lowest minimum as the algorithm will tend to fall into the local minimum closest to the starting point. On the other extreme, when over-constrained the algorithm fails to find a minimum and usually diverges. In principle choosing a number of constraints equal to the number of varying parameters should ensure proper behavior of the minimization function. However, in all but the simplest cases both parameters and constraints can be correlated within their own spaces and it becomes very difficult to track the real dimensionality of the two spaces and make them more or less equal. A good example of such correlation could be the length and pole tip field of a quadrupole magnet, both of which directly affect the focusing length in almost the same way. Add to this the possibility to adjust the tolerance of each fitted parameter, and very quickly the best method to find a particular solution mostly relies on trial-and-error combined with experience. Having a tool such as the program described in this paper which allows users to quickly change the starting point and try various fitting combinations becomes a great asset.

\section{Description of the user interface}
Although the goal of this article is not a comprehensive description of the program (a manual is available for that purpose), a short description of the key features is necessary to illustrate its capabilities and the advantages of using this tool for optics design and optimization. The subsequent sections will follow the steps taken by  a typical user in designing and entering a particular beam line, study its behavior and optimize some of its parameters for a particular purpose.

\subsection{Entering a design}
\begin{figure}
\begin{center}
\includegraphics[scale=0.8]{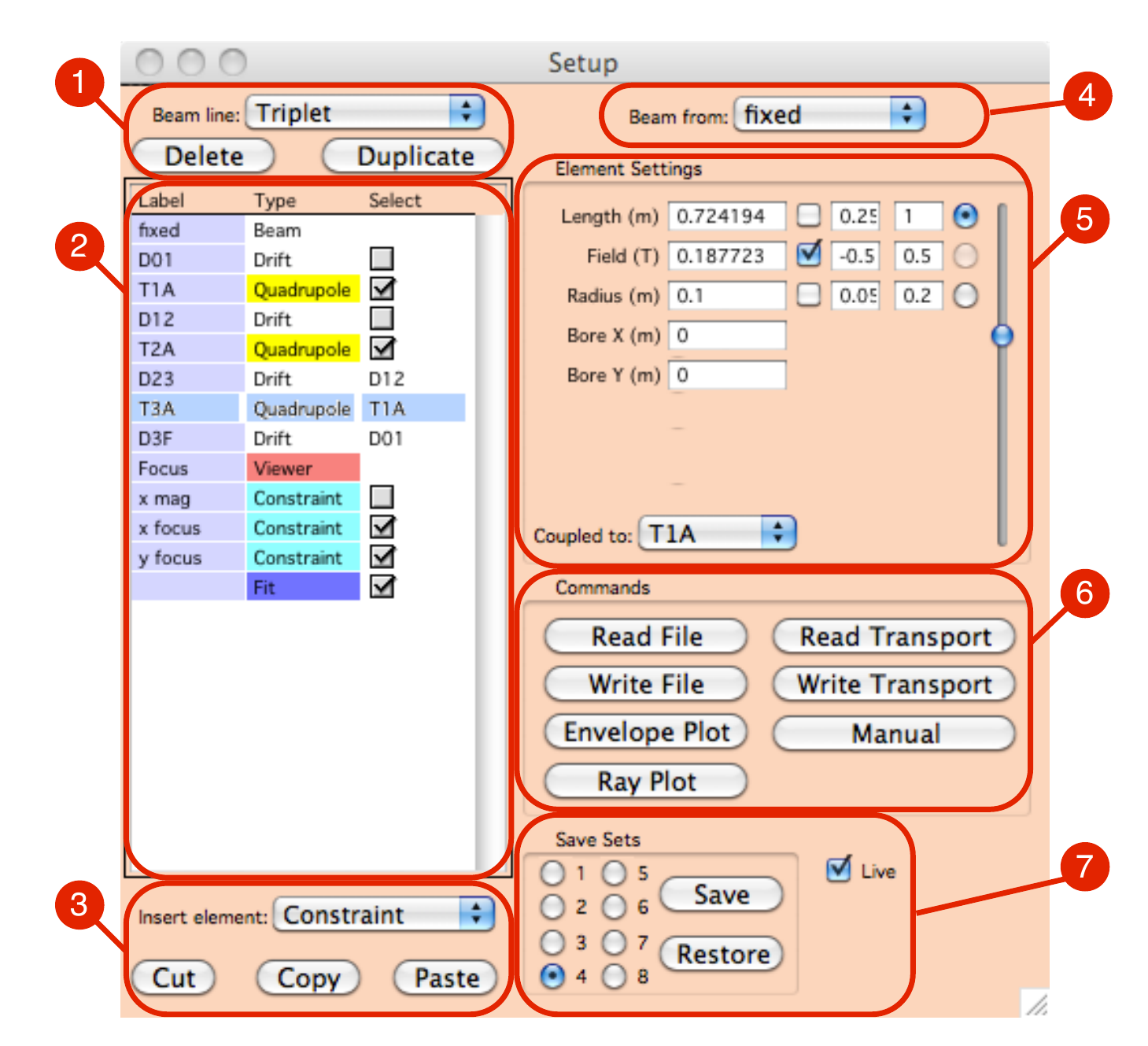}
\caption{Main setup panel of Tuner. Each functionality included in the panel is labeled with circled numbers. \ding{202} Beam line manipulation allows to select, as well as duplicate or delete the curent beam line. \ding{203} List box showing the configuration of the current beam line, including the label, type and selection of each element. One or more contiguous elements can be selected. \ding{204} Editor actions on the selected element(s). The insertion or pasting always occurs downstream of the last selected element. \ding{205} Selection of the incoming beam, either fixed or from another beam line. \ding{206} Details of the selected element. \ding{207} General commands. \ding{208} Save set manipulation.}
\label{MainGUI}
\end{center}
\end{figure}

Figure \ref{MainGUI} shows the main setup panel used to manage beam line designs. It is possible to enter a design either manually using the simple editing tools provided in the panel (\ding{204}), or by importing an already existing Transport file (\ding{207}). In this latter case, the program will parse the file and create the appropriate elements in the beam line. Some additional editing might be necessary depending on the information contained in the original Transport file.
The beam line configuration currently selected as indicated in \ding{202} is displayed in the list box \ding{203}, which lists the various elements starting from the beam definition. The detailed parameters of each element are displayed in \ding{206}. Note that several contiguous elements can be selected and used with the editing tools. This facilitates the entering of configurations composed of identical sections, as is often the case. The "Live" button in \ding{208} can be used to disable the automatic updating of the envelope or ray plots while entering or adjusting the parameters of the various elements, which can cause calculation errors when they are not defined properly. The alternative method is to close all windows displaying dynamically the current configuration.

Apart from the standard elements found in beam lines, the program Tuner introduces "viewer" and "fit" elements that can be placed anywhere within the configuration. A viewer element is used to visualize the transfer, beam  and inverse matrices as well as the beam ellipses and their parameters. The fit elements break the configuration into subsections that can be fitted separately. It is usually placed after a set of constraints that define the fitting criteria, although any enabled constraint located between two enabled fit elements is taken into the minimization function as well. 

\subsection{Studying behavior}
\begin{figure}
\begin{center}
\includegraphics[scale=.35]{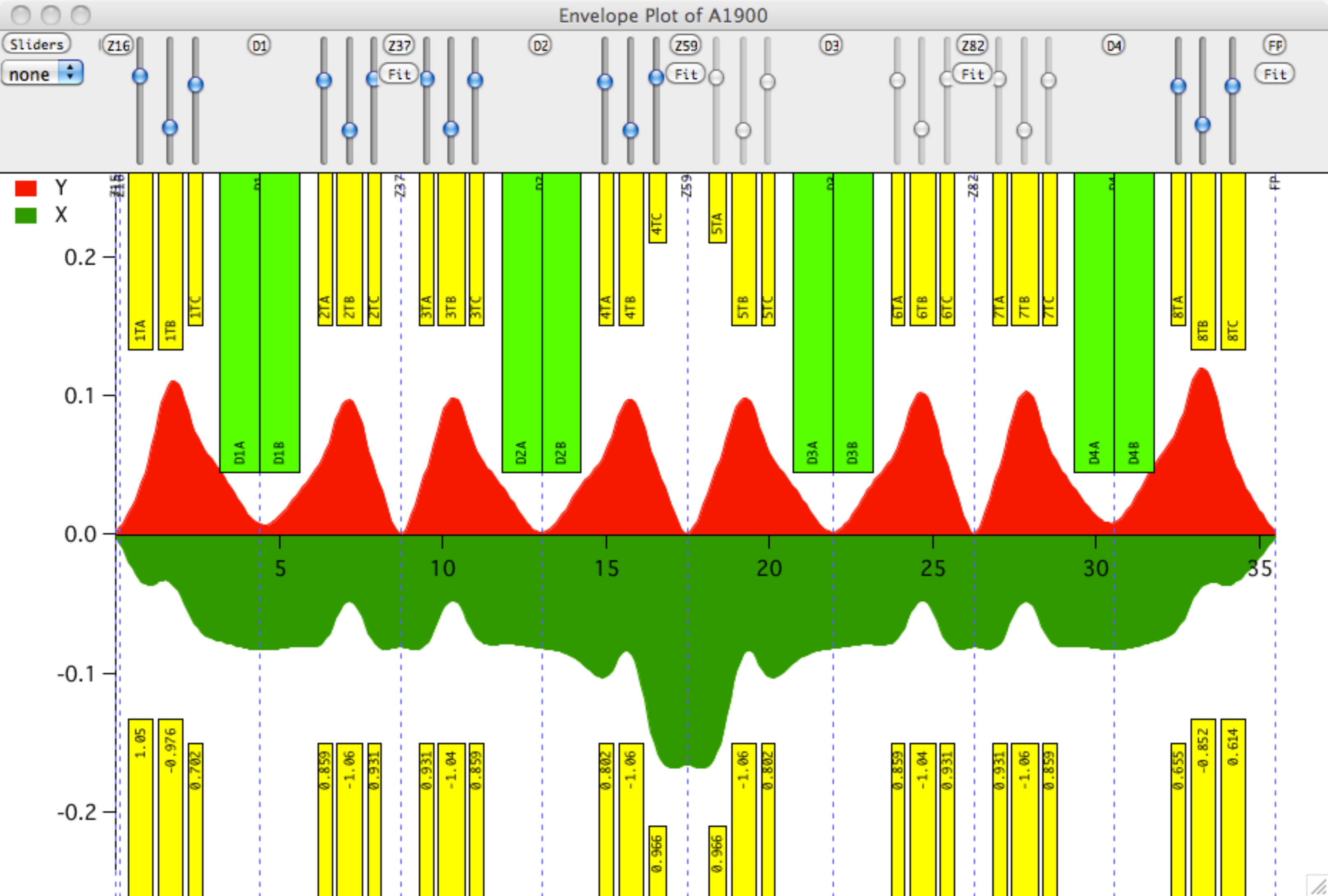}
\caption{Envelope plot of the A1900 fragment separator in its standard optics mode, after import from the Transport file. The top panel can be opened or closed by pressing on the "Sliders" button. Lens elements that are selected in the list of elements can then be manually adjusted using individual sliders. Coupled lenses are indicated by disabled sliders. The viewers are labeled using the labels entered in the list of elements, and are used to open viewer display windows. The fit buttons appear whenever the fit elements are selected in the list of elements, and trigger the fitting procedure when pressed. The input emittance used in this calculation corresponds to the standard A1900 acceptance: $\Delta$a = $\pm$ 50 mrad, $\Delta$b = $\pm$ 40 mrad, and $\Delta$d = $\pm$ 3 \%.}
\label{A1900}
\end{center}
\end{figure}

Figure \ref{A1900} shows the envelope plot obtained from reading the standard Transport file of the A1900 fragment separator \cite{A1900}, and adding the fit elements at the appropriate locations (a viewer element is automatically created for each drift found in the Transport file with its length set to zero). Figure \ref{A1900} also shows the top panel located above the envelope plot, that is used to control lens elements (quadrupoles or solenoids), open viewer displays or trigger fitting. 
Sliders corresponding to lens elements appear in the top panel whenever they are selected in the element list (see \ding{203} in Fig. \ref{MainGUI}). They can be used to manually adjust each selected lens to alter the optics of the beam line. The main purpose of this functionality is to enable an easy manual exploration of optical configurations. As the sliders are moved, the envelope and ray plots are updated, as are any viewer displays currently open. The variation range of the sliders is taken from the variation limits imposed on the strength of the lens elements (see section \ref{varlim}). Note in the figure that the sliders for quadrupole triplets 5, 6 and 7 are grayed out, indicating that these lenses are coupled to others located upstream. The standard A1900 optics uses a symmetric configuration for all quadrupole triplets but the first and last ones, due to the shorter length on the production target side. The symmetry around the dispersive focal plane at viewer Z59 is readily visible on Fig. \ref{A1900}. As the strength is moved on a lens, any other that is coupled to it tracks its value simultaneously.

\begin{figure}
\begin{center}
\includegraphics[scale=.7]{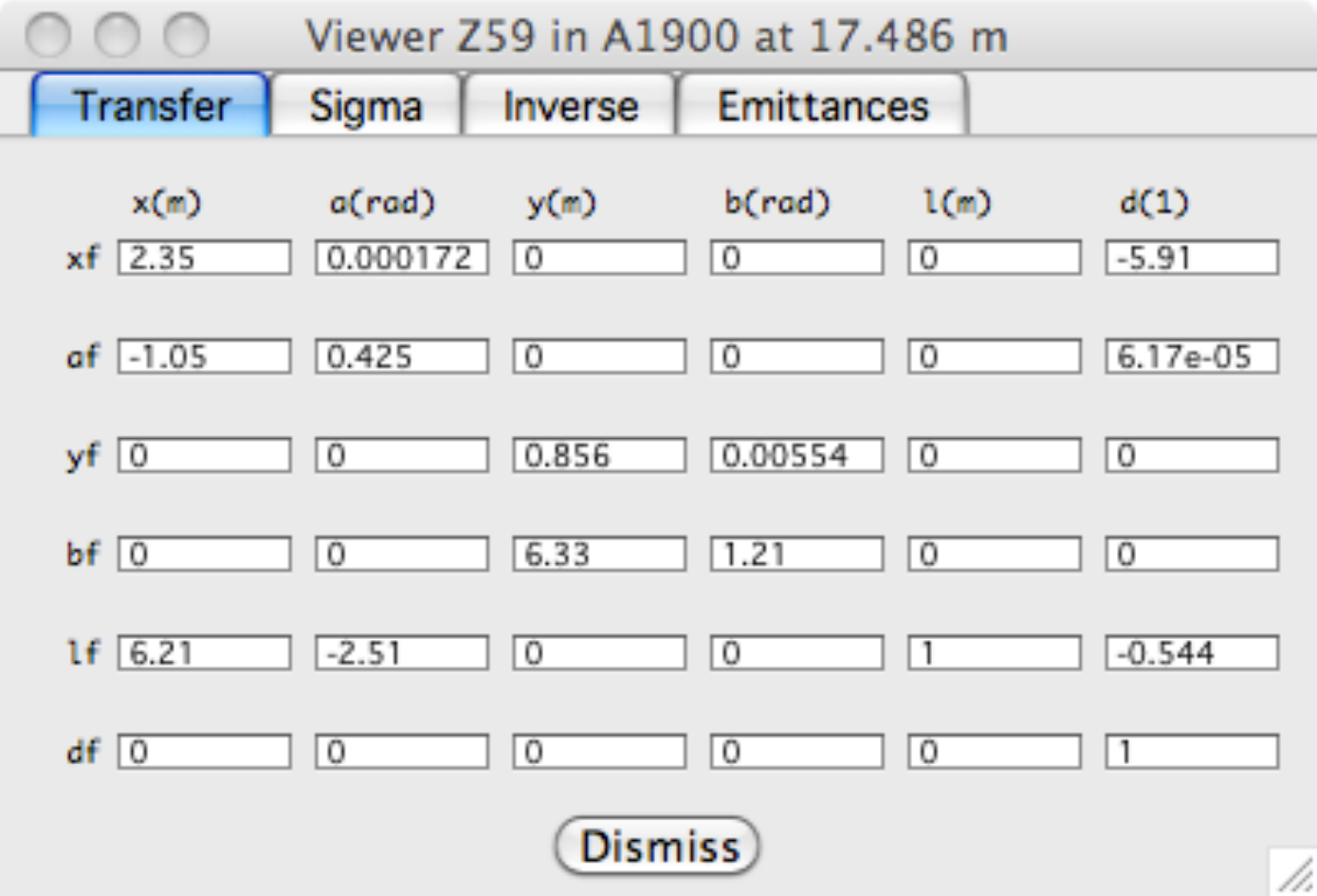}
\caption{Transfer matrix of the A1900 fragment separator at the dispersive image Z59 in standard optics mode. The units are meters and radians for the horizontal and vertical sizes and angles respectively, meters for the trajectory length and a number normalized to unity for the momentum deviation. See text for details.}
\label{A1900Matrices}
\end{center}
\end{figure}

While adjusting individual lenses, it is often helpful to watch the evolution of the transfer or beam matrices. This is easily done by opening any of the viewer displays. An example of a viewer display of the transfer matrix at the dispersive image (Z59) is shown in Fig. \ref{A1900Matrices}. This matrix shows the double focus conditions  as well as the dispersion of -5.91 cm/\% at this location. The numbers are dynamically updated as the optical configuration is being altered.
The viewer displays also feature tabs to view the sigma and inverse matrices, as well as plots of the horizontal and vertical emittances, together with their associated parameters. In addition, a menu located at the top left corner of the envelope plot top panel can be used to plot matrix elements along the beam line. Although only one matrix element can be displayed at a time, a direct visualization of its evolution as optical elements in the beam line are being adjusted is often a better guide than the numbers displayed in the matrix.

\subsection{Calculation of a new optics configuration}
As an illustration of the possibilities of this program, one could imagine tuning the A1900 fragment separator in a configuration in which the horizontal and vertical angular sizes of the beam can be constrained using slits at locations Z59 and Z37. The standard configuration cannot be used for this purpose because both Z37 and Z59 are dispersive images where focussing conditions are realized in both horizontal and vertical planes, and momentum dispersion is non-zero. The procedure used to achieve a particular optical configuration involves an iterative process in which the user goes back and forth between manual adjustment of the lenses, and fitting of their strengths to fulfill a number of constraints. In the particular case of an angle-cutting configuration, the goal is to configure the optics to leave only the angle dependent position matrix elements (y/b) and (x/a) non-zero at Z37 and Z59, respectively. When these conditions are realized, the horizontal and vertical positions are determined by the respective angles at the object, and any position cut translates directly into a cut in angle. The constraints used for this particular configuration are listed in Table \ref{constraints}.
\begin{table}[htdp]
\begin{center}
\begin{tabular}{|c|c|c|} \hline
Constrain & Location & Comment \\ \hline
(y/b)=0 & D1 & Vertical image  \\
(a/a)=0 & D1 & Horizontal parallel \\ \hline
(y/y)=0 & Z37 & Vertical magnification = 0 \\
(y/b)=-1 & Z37 & Vertical angle cutting coefficient \\
(x/d)=0 & Z37 & Horizontal dispersion = 0 \\ \hline
(x/x)=0 & Z59 & Horizontal magnification = 0 \\
(x/a)=-1 & Z59 & Horizontal angle cutting coefficient \\
(b/b)=0 & Z59 & Vertical parallel \\ \hline
\end{tabular} 
\caption{List of constraints used in the optimization of the angle cutting configuration calculated for the first two sections of the A1900 fragment separator. The locations correspond to labeled viewers along the beam line. The vertical and horizontal angle cutting slits are located at Z37 and Z59, respectively, with a coefficient of -1 mm/mrad. The horizontal dispersion is set to 0 at Z37, and the third triplet is a mirror image of the second, in order to achieve a doubly achromatic condition at Z59 (both (x/d) and (a/d) equal to 0).}
\label{constraints}
\end{center}
\end{table}
The constraints located in the middle of the first dipole D1 are meant to control the envelope in both horizontal and vertical planes, in particular with regards to the vertical gap. The vertical angle cutting condition is realized at Z37 by setting (y/y) to 0, and fixing the angle cutting coefficient (y/b) to -1 mm/mrad. This choice is of course arbitrary, and other solutions with a different coefficient are possible. The horizontal dispersion (x/d) is constrained to 0 in order to realize a doubly achromatic condition at Z59 (both (x/d) and (a/d) equal to 0). Together with the third triplet which is a mirror image of the second, this constraint effectively cancels the dispersions after the second dipole D2. This is necessary in order to setup good conditions for the horizontal angle cutting at Z59, for which no horizontal dispersion should be present.

There are six quadrupole lenses available to adjust the five constraints between the object and the Z37 location, therefore several solutions are possible, making the fitting highly dependent on the starting conditions. However, not all solutions may be acceptable, in particular with regards to the angular and momentum acceptances. This is where the manual adjustment of lenses can be efficiently guided by the direct feedback provided by the program. It is possible to approach an approximate solution in this way, that is suitable as a starting point for the fitting of the required constraints. Using the save sets, the user can easily store and recall configurations before and after fitting in case the fit failed or found a minimum that is not satisfactory. Different starting conditions can be stored in different save sets. As it is both an exploratory and iterative process, different ideas can easily be attempted to achieve the configuration goals. As the number of iterations is limited to 100 in the minimization algorithm, it is possible, in particular when fitting in large dimensional space such as the 6 quadrupole strengths in this example, that the fitting stops before reaching a minimum. The minimization might also stop if no gradient is found, indicating that a local minimum was reached. To escape this type of situation, a noise parameter is provided that will nudge the starting values before attempting the next minimization. This parameter as well as the value of the fitting function are displayed in the fit element details, therefore the user can quickly assess the quality of the fit and restart the minimization until it reaches a satisfactory minimum.

\begin{figure}
\begin{center}
\includegraphics[scale=1]{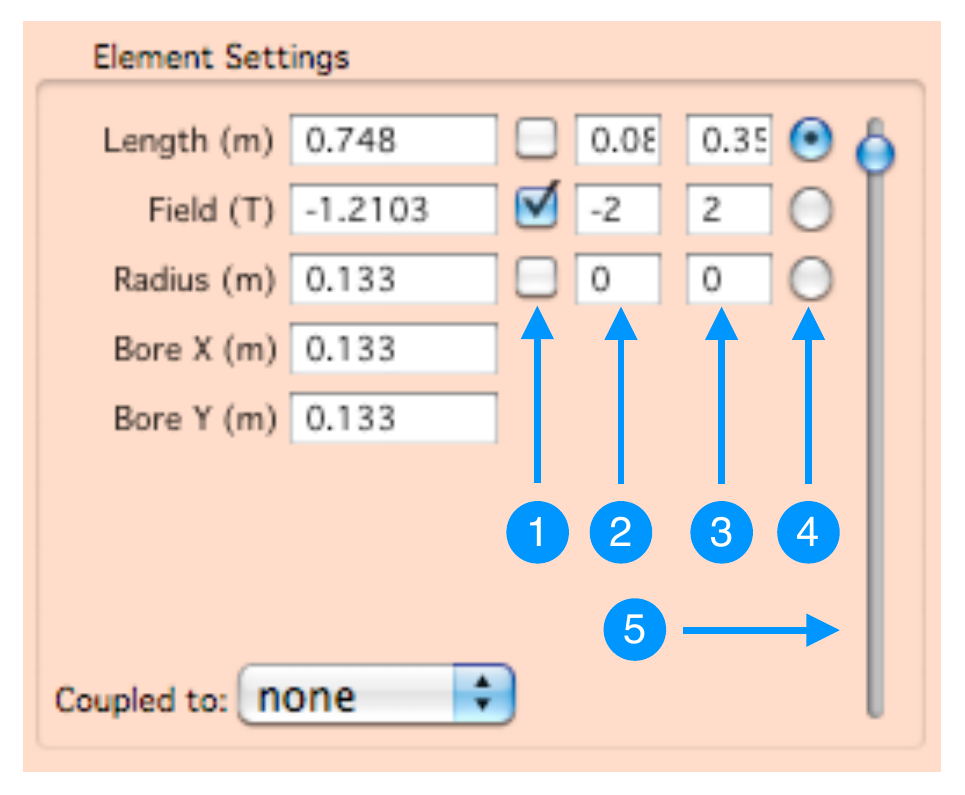}
\caption{Example of element setting controls embedded in the main GUI (see \ding{206} in Fig. \ref{MainGUI}). Checkboxes \ding{202} are used to select which parameter(s) are varied by the minimization algorithm. They also control which parameter(s) are varying when the selected element is coupled to an other one upstream. Entries \ding{203} and \ding{204} are the lower and upper limits for each parameter, used both as soft boundaries by the fitting algorithm, and as limits for the sliders. The radio buttons \ding{205} are used to select the parameter controlled by the slider \ding{206}, allowing parameters other than the strength in lenses to be adjusted graphically.}
\label{ElementSetting}
\end{center}
\end{figure}

The control of which parameters of optical elements are allowed to vary during the fitting procedure is illustrated in Fig. \ref{ElementSetting}, for the case of a quadrupole lens. 
Note that the parameters that can be adjusted are not limited to the strength of lenses, which is the usual choice when exploring optical possibilities of an existing beam line, but to others as well. With this capability this program can be used in the design phase of ion optical systems as well, and options or limitations related to the sizes and dimensions of elements used in the design can be explored in a very straightforward way.

\begin{figure}
\begin{center}
\includegraphics[scale=.35]{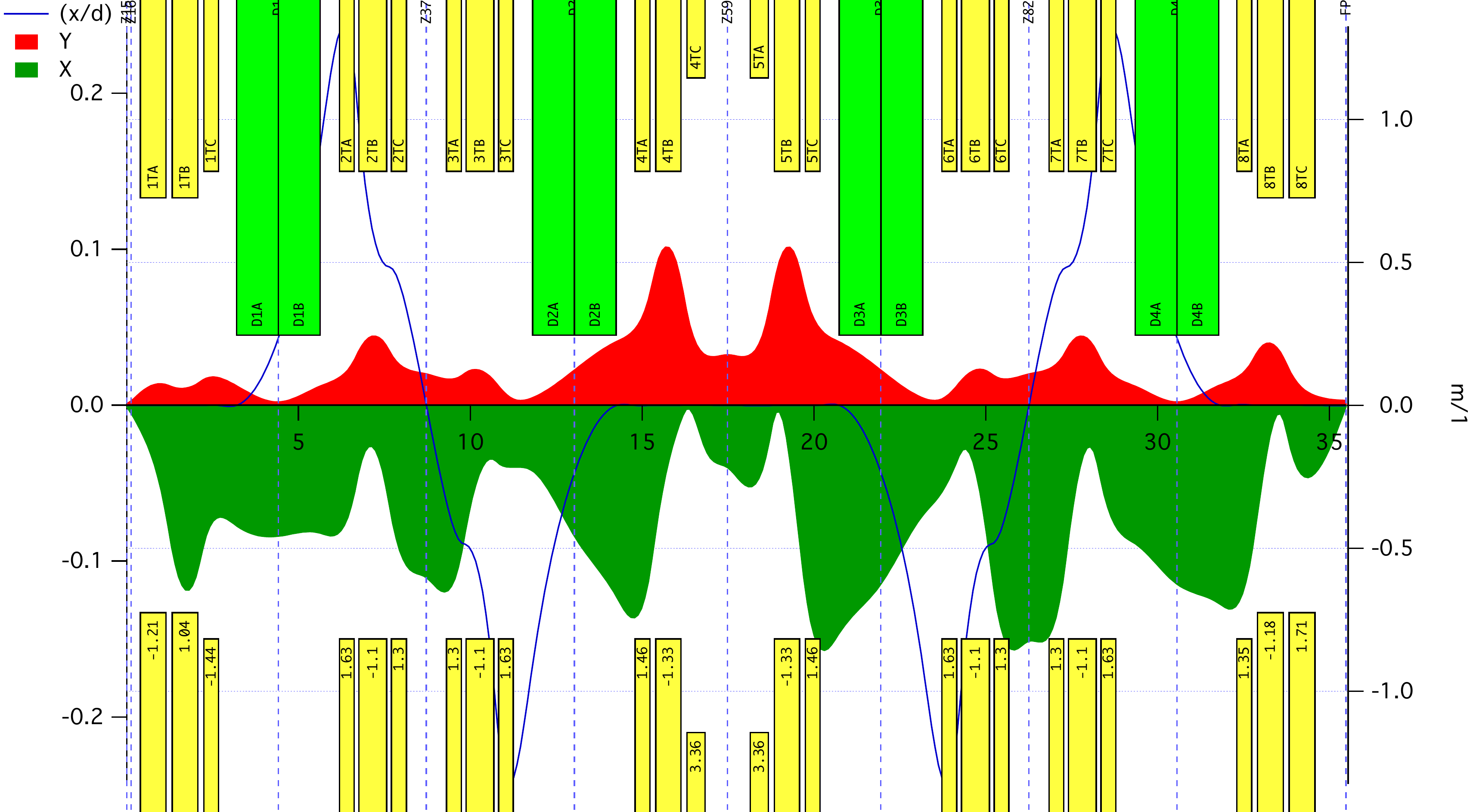}
\caption{Envelope plot of the A1900 fragment separator tuned to the angle cutting configuration described in the text. The input emittance used in this graph is slightly smaller than the standard A1900 acceptance: $\Delta$a = $\pm$ 40 mrad, $\Delta$b = $\pm$ 20 mrad, and $\Delta$d = $\pm$ 3 \%.}
\label{A1900Angle}
\end{center}
\end{figure}

The envelope plot of a possible angle cutting configuration of the A1900 fragment separator, based on the constraints listed in Table \ref{constraints} and the symmetries built into the configuration, is displayed in Fig. \ref{A1900Angle}. Also drawn is the horizontal dispersion (x/d) along the beam line, its scale being displayed on the right hand side of the plot. The mirror symmetry of the A1900 is used from the Z59 location to set the remaining quadrupole triplets, with the exception of the last one for which a slightly different tune is calculated to achieve double focussing at the focal plane FP.
The first order transfer matrices calculated at Z37 and Z59 for this configuration are 
$$R_{Z37}=\left[\begin{array}{cccccc}-1.09 & -2.76 & 0 & 0 & 0 & 2.3\times10^{-7} \\0.22 & -0.361 & 0 & 0 & 0 & -0.579 \\0 & 0 & 1.32 \times10^{-5} & -1 & 0 & 0 \\0 & 0 & 1 & 0.28 & 0 & 0 \\-0.629 & -1.6 & 0 & 0 & 1 & -0.242 \\0 & 0 & 0 & 0 & 0 & 1\end{array}\right]$$
and
$$R_{Z59}=\left[\begin{array}{cccccc}-7.48\times10^{-7} & -1 & 0 & 0 & 0 & -3.28\times10^{-6} \\1 & -0.468 & 0 & 0 & 0 & -6.46\times10^{-4} \\0 & 0 & 1.03 & 1.61 & 0 & 0 \\0 & 0 & -0.619 & -7.43\times10^{-8} & 0 & 0 \\-2.95\times10^{-6} & -6.44\times10^{-4} & 0 & 0 & 1 & -0.242 \\0 & 0 & 0 & 0 & 0 & 1\end{array}\right]$$
respectively.
This type of optics could be used for instance in an attempt to produce secondary beams with a large proportion of spin polarization, by selecting a particular region of the transverse momentum distribution produced by the projectile fragmentation process. Earlier studies have shown that significant spin polarization can be observed on the tails of these distributions \cite{Groh}.

\section{Conclusions and prospects}
This paper presents a short overview of the program Tuner, a new computing tool aimed at facilitating and accelerating the design and optimization of ion optical systems. This program calculates first order optics of beam line configurations dynamically, taking advantage of present-day computing speed and graphical interface frameworks. Its main advantage is the possibility to quickly explore the usually vast parameter space typically found in ion optical systems, and select the best design or tuning parameters for achieving its goals. This can be achieved thanks to the ability to easily select a particular starting point before letting the optimization algorithm search for the closest minimum corresponding to a set of constraints.

Although the primary use of this program is to explore configurations of existing ion optical systems, it can also be used as a design tool to optimize the physical parameters of conceptual systems, a first and necessary step before full simulation programs can be used to further characterize the performances and limitations. At the other end of the user spectrum, this program can also be an efficient educational tool for students or individuals learning beam optics. As a complementary material to textbooks and courses, the program Tuner allows users to experience directly the effect of beam line elements on the optics of a system, giving them a more intuitive feel on how to tune them correctly and achieve particular configurations. One could even envision using this tool for problem assignments in a class.

The features demonstrated in this paper correspond to the version available at the time of this writing, however further improvements and future additional features are envisioned. Probably the most useful feature would be the addition of second order effects that can be calculated analytically. Together with the addition of sextupoles, this improvement would enable more realistic calculations of the envelopes, as well as the possibility to minimize the effects of these aberrations. Other important characteristics of ion optical systems are their acceptances. Even though the program Tuner gives an overall measure of the acceptances by tracing the beam envelope along the beam line and showing where it intercepts any of the physical limits of the elements, a more rigorous approach should use a random simulation of rays, keeping only those that can pass through the system. This type of Monte-Carlo simulation can easily be implemented in Tuner using the Igor Pro$\textsuperscript{\textregistered}$ framework. Finally, this tool could be used as an online visualization of the optics of a running device, using for instance the widely used EPICS protocol to communicate real field values from the magnets to the program. An EPICS driver for Igor Pro$\textsuperscript{\textregistered}$ is already available for that purpose. A set of theory-to-reality ratios could be used to normalize the calculation to a known tune of the device, in order to compensate for inaccuracies in the physical definition of the beam line, as well as effects not included in the calculation.

The program Tuner is available from the following web page: \href{http://www.nscl.msu.edu/~bazin//Tuner}{www.nscl.msu.edu/$\sim$bazin/Tuner}. The Tuner package contains a data file containing the predefined beam line configurations shown in this article, and that can be used as examples, as well as a detailed manual and the Igor Pro$\textsuperscript{\textregistered}$ procedure files themselves. Keeping these procedure files separate from the data files ensures that any update of the code is carried over to all data files. As with any kind of software, bugs and unforeseen error conditions always happen, and it is the hope of the author that users can report their findings and voice their suggestions so that the program can improve over time. Users that want to be notified of new versions or updates should send a brief Email to the author.

This work was supported by the US National Science Foundation under grant number PHY-0606007.



\end{document}